\documentclass[referee]{aa}
\usepackage{amsmath,mathrsfs}
\usepackage{natbib}
\usepackage{graphicx}
\begin{document}
\headnote{Research Note}
\title{Scattering of Polarized Radiation by Atoms in Magnetic
and Electric Fields}
\author{Yee Yee Oo\inst{1}
\and K. N. Nagendra\inst{2}
\and Sharath Ananthamurthy\inst{1}
\and G. Ramachandran\inst{2} }
\institute{ Department of Physics, Bangalore University,
Bangalore-560 056, India \and
Indian Institute of Astrophysics, Bangalore-560 034, India }
\date{Received dd mm yy  / Accepted dd mm yy}

\abstract{
The polarization of radiation by scattering on an atom embedded in 
combined external quadrupole electric and uniform magnetic fields is
studied theoretically. Analytic formulae are derived for the scattering
phase matrix. Limiting cases of scattering under Zeeman effect, and 
Hanle effect in weak magnetic fields are discussed.
 }

\titlerunning{ Scattering }
\authorrunning{ Oo et al. }

\maketitle
\section{Introduction}
Scattering of polarized radiation by an atom is a topic of considerable
interest to astrophysics, especially with the advent of imaging polarimeter
systems like ZIMPOL I \& II~\citep{zimpol}
which reach accuracies of the order of $10^{-5}$ for measuring the Stokes
parameters characterising the observed radiation. The concept of scattering
phase matrix connecting the Stokes vector ${\boldsymbol{\cal S}}'$
of incident radiation to the Stokes vector ${\boldsymbol{\cal S}}$
of scattered radiation was introduced quite early in the context of Rayleigh
scattering.
Landi Degl'Innocenti~\citep{landi-1, landi-2, landi-3, landi-4},
Landi Degl'Innocenti, Bommier and Sahal-Br$\acute{\rm e}$chot~\citep{lbs1, lbs2, lbs3}
and Bommier~\citep{bom-1, bom-2} developed a comprehensive theoretical framework
to describe the generation and transfer of polarized radiation in spectral lines,
formed in the presence of an external magnetic field.
In the context of radiation transfer work, Stenflo and Stenholm~\citep{stenflo2} 
and Rees~\citep{ree},
used complete frequency redistribution (CRD) in the resonance scattering, 
and Dumont et al.~\citep{dumont}, Rees and Saliba~\citep{ree2},
Nagendra~\citep{nagendra1, nagendra2}, Faurobert~\citep{faur},
Ivanov et al.~\citep{ivan} and later works employed partial frequency redistribution
(PRD) line scattering mechanisms in the absence of magnetic field. 
The Hanle effect is a depolarizing phenomenon which arises due to `partially 
overlapping' magnetic substates in the presence of weak magnetic fields, where
the splitting produced is of the same order as or less than the natural widths.
Favati \textit{et al.}~\cite{favati} proposed 
the name `second Hanle effect' for a similar effect in `electrostatic fields'.
Casini and Landi Degl'Innocenti~\citep{casini} have discussed
the problem in the presence of electric and magnetic fields for the particular case
of hydrogen lines. The relative contributions of static external electric
fields, motional electric fields and magnetic fields in the case of hydrogen
Balmer lines, have been studied
by Brillant \textit{et al.}~\citep{brillant}. A historical perspective and extensive
references to earlier literature on polarized line scattering can be found in
Stenflo~\citep{stenflo-94}, Trujillo Bueno \textit{et al.}~\citep{tru} and 
Landi Degl'Innocenti and Landolfi~\citep{ll}.
The purpose of this paper is to derive the scattering phase matrix for the case of
combined magnetic and electric quadrupole fields with arbitrary strengths. 
The particular case of transitions between $J=0$ and $J=1$ states is considered
following Oo \textit{et al.}(2004; 2005). 

\section{Theoretical Formalism}
\begin{figure}
\begin{center}
\includegraphics[width=6cm,height=6.5cm]{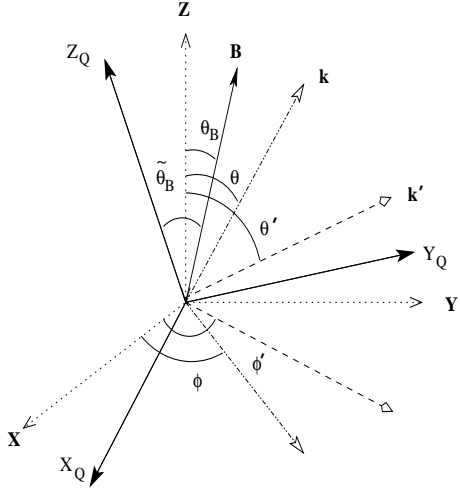}
\end{center}
\caption{ The scattering geometry:
$({\rm X}_{Q}, {\rm Y}_{Q}, {\rm Z}_{Q})$ refers to the Principal
Axes Frame (PAF) characterising the electric quadrupole field.
The radiation is incident along $(\theta', \phi')$ and scattered along
$(\theta, \phi)$ with respect to the astrophysical reference frame
denoted by $({\rm X}, {\rm Y}, {\rm Z})$. The magnetic field
${\vec B}$ is oriented along $(\widetilde \theta_{B}, \widetilde \phi_{B})$
with reference to PAF and $(\theta_{B}, \phi_{B})$ with reference to the
astrophysical reference frame (the azimuthal angles $\widetilde \phi_{B}$
and $\phi_{B}$ are not marked in the figure). \label{PAF}}
\end{figure}

The Hamiltonian for an atom, when it is exposed to an external magnetic field
${\vec B}$ together with an arbitrary external Coulomb field $\Phi$, is of the 
form~\citep{Oo}
\begin{equation}
H = H_{0} + g\, {\vec J} \cdot {\vec B} + \sum^{\infty}_{l=0}
\sum^{l}_{m=-l} (-1)^{m}\, V_{lm}\, Q_{l-m} \ ,
\label{hint}
\end{equation}
where $g$ denotes the gyromagnetic ratio, ${\vec J}$ the
total angular momentum operator for the atom with components $J_{x}, J_{y},
J_{z}$ while $V_{lm}$ denote the $2^{l}$-pole components of $\Phi$ and 
the components $Q_{l-m}$ characterise the electric charge distribution
inside the atom. It is well-known that the magnetic field ${\vec B}$ splits an
eigenstate of $H_{0}$ with energy $E$ and total angular momentum quantum number 
$J$ into $(2J+1)$ equally spaced levels $|J\, M\rangle$ with energies
$E_{M}=E+gBM$, where $B=|{\vec B}|$ and $M$ denotes the magnetic
quantum number with respect to an axis of quantization chosen along ${\vec B}$.
If the atomic states are eigenstates of parity, the $l=1$ term makes no
contribution. In an external electric quadrupole field with $l=2$, 
$V_{2m}$ may equivalently
be expressed in terms of the cartesian components $V_{\alpha,\beta}$ with
$\alpha, \beta = x, y, z$ of a traceless symmetric second rank tensor, which
defines its own Principal Axes Frame (PAF), wherein $V_{\alpha,\beta}=V_{\alpha,
\alpha}\, \delta_{\alpha,\beta}$, so that
\begin{equation}
\sum^{2}_{m=-2} (-1)^{m}\, V_{2m}\, Q_{2-m} = A\, \Bigl[ 3J_{z}^{2} - {\vec J} 
\cdot {\vec J} + \eta\, \{J_{x}^{2} - J_{y}^{2} \} \Bigr]\ ,
\end{equation}
where $A=Q\, V_{zz}/\{4J(J-1)\}$, if $Q$ denotes the electric quadrupole
moment of the atom and $\eta = (V_{xx} - V_{yy})/V_{zz}$ denotes the asymmetry 
parameter of the field. 
In such a case, the $(2J+1)$ substates $|\psi_{k}\rangle$ where $k=1,2,\cdots,(2J+1)$ 
with energies $E_{k}$ are neither equally spaced nor are they identifiable as 
$|J\, M\rangle$ states.
We may, however, represent them as 
\begin{equation}
|\psi_{k}\rangle = \sum^{J}_{M=-J} a^{k}_{M}({\vec B}, A, \eta)\, |J\, M\rangle \ ,
\end{equation} 
in terms of $|J\, M\rangle$ states defined with
respect to PAF, where the expansion coefficients $a^{k}_{M}$ as well as the 
energies $E_{k}$ are not only functions of $B, A, \eta$ but also of the angles
$(\widetilde \theta_{B}, \widetilde \phi_{B})$ of ${\vec B}$ with respect to
PAF~(see Fig.~\ref{PAF}). 
For a detailed discussion for $J=1, \frac{3}{2}$ (see Oo et al. 2004; 2005).

The PAF itself may, in general, be different from the astrophysical frame,
in which case  
\begin{align}
|\psi_{k}\rangle &= \sum^{J}_{m=-J} c^{k}_{m} \, |J\, m\rangle \ ; &
k &= 1,2,\cdots,(2J+1)\ , \label{ast}
\end{align}
in terms of the $|J\, m\rangle$ states defined with respect to the astrophysical
frame and 
\begin{equation}
c^{k}_{m}=\sum^{J}_{M=-J} a^{k}_{M}({\vec B}, A, \eta)\, D^{J}_{mM}
(\alpha_{Q}, \beta_{Q}, \gamma_{Q})\ , \label{abg}
\end{equation} 
if $(\alpha_{Q}, \beta_{Q}, \gamma_{Q})$ denote the Euler angles of the PAF
with respect to the astrophysical frame. If the magnetic field alone is
present, $|\psi_{k}\rangle$ are identical with $|J\, M\rangle$ states and
$c^{M}_{m}=D^{J}_{mM}(\phi_{B},\theta_{B},0)$ in Eq.~(\ref{ast}). 

We now consider the scattering of polarized radiation by an atom 
which makes a transition from an initial state $|\psi_i\rangle$ with
energy $E_{i}$, total angular momentum $J_{i}$ and parity $\pi_{i}$
to a final state $|\psi_f\rangle$ with energy $E_{f}$, total angular
momentum $J_{f}$ and parity $\pi_{f}$ when polarized radiation with 
frequency $\nu'$ is incident on the atom in the astrophysical frame
in a direction $(\theta',\phi')$ and gets scattered into a direction
$(\theta,\phi)$ with frequrncy $\nu$. The left and right circular
states of polarization as defined by Rose~\citep{rose} are denoted by
$\mu=\pm 1$. The second order transition matrix element for scattering
of polarized radiation may then be written, with respect to the polarization
states as 
\begin{equation}
T_{\mu \mu'} = \sum_{n} {\cal E}_{fn}(\mu)\, \Phi_{n}\, 
{\cal A}_{ni}(\mu') \ , \label{tmumu}
\end{equation}
where the summation is with respect to the intermediate states 
$|\psi_n\rangle$ of the atom with energy $E_{n}$, total angular momentum 
$J_{n}$ and parity $\pi_{n}$.
Following~\citep{Oo}, the matrix elements for emission from
$|J_{u}\, m_{u}\rangle$ to $|J_{l}\, m_{l}\rangle$ are of the form
\begin{eqnarray}
{\cal E}_{m_{l}m_{u}}(\mu)&=& \sum_{L} {\cal J}_{L
}\,
\sum^{L}_{M_{L}=-L} C(J_{l},L,J_{u};m_{l},M_{L},m_{u})\, \nonumber \\
& & D^{L}_{M_{L}\mu}(\phi,\theta,0)^{*}\ , \label{tme}
\end{eqnarray}
where ${\cal J}_{L}$ is proportional to the reduced matrix element.
The complex conjugate of Eq.~(\ref{tme}) defines the matrix elements
${\cal A}_{m_{u}m_{l}}(\mu)$ for absorption of radiation with
polarization $\mu$ incident along $(\theta,\phi)$ leading to
$|J_{u}\, m_{u}\rangle$ from $|J_{l}\, m_{l}\rangle$.
Using the notations $\omega=2\pi \nu;\ \ \omega'=2\pi \nu'$;\ \ $\omega_{n}
=E_{n}-E_{f}$ and $\omega'_{n}=E_{n}-E_{i}$, the profile function
$\Phi_n= (\omega_{n}-\omega-{\rm i}\Gamma_{n})^{-1}=
(\omega'_{n}-\omega'-{\rm i}\Gamma_{n})^{-1}$ where
the width associated with $|\psi_n\rangle$ is denoted by $\Gamma_{n}$ and 
energy conservation requires $E_{i}+\omega'=E_{f}+\omega$.
Angular momentum and parity are conserved individually during the absorption
and the emission.  

\section{Scattering Phase Matrix}
If ${\boldsymbol{\cal S}}'$ denotes the Stokes vector,
which characterizes the state of polarization of the incident radiation, 
the Stokes vector ${\boldsymbol{\cal S}}$ characterizing the scattered 
radiation is
\begin{equation}
{\boldsymbol{\cal S}}= {\cal R}\ \ {\boldsymbol{\cal S}'}\ ,
\end{equation}
where ${\cal R}$ is a $(4 \times 4)$ matrix whose elements are of the form 
\begin{equation}
{\cal R}_{pp'}= \sum_{nn'} \Phi_{n}\, \Phi^{*}_{n'}\,
{\cal P}^{nn'}_{pp'}\ ,
\label{rematrix}
\end{equation}
where the phase matrix elements are given by
\begin{eqnarray}
& &{\cal P}^{nn'}_{pp'}= \frac{1}{2}
\sum_{\mu, \mu' \atop \mu'', \mu'''}{(\sigma^{\gamma}_{p'})}_{\mu'\mu'''}\,
{(\sigma^{\gamma}_{p})}_{\mu''\mu}\,
\nonumber \\
& & Tr \Bigl[ {\cal E}(\mu)\,
{\cal C}^{n}\,
{\cal A}(\mu')\, {\cal C}^{i}\,
{\cal A}^{\dag}(\mu''')\, {\cal C}^{n'}\, {\cal E}^{\dag}(\mu'')\,
{\cal C}^{f} \Bigr] \ ,\label{spm}
\end{eqnarray}
with $p,p'=0,1,2,3$, using the density matrix formalism~\citep{mcmaster} for
polarization of radiation. 
The $\sigma^{\gamma}_{p}$ with $p=1,2,3$ are Pauli matrices
with respect to basis states $\mu=\pm 1$ for radiation, while $\sigma^{\gamma}_{
0}=1$. We use
the notation $Tr (\equiv \sum_{m_f})$ to denote the Trace of
the $(2J_f+1) \times (2J_f+1)$ matrix contained within the square
bracket. 
The ${\cal C}^{i}, {\cal C}^{f}, {\cal C}^{n}$
and ${\cal C}^{n'}$
are defined in terms of their elements
\begin{align}
{\cal C}^{i}_{m_{i}m'_{i}}&= c^{i}_{m_{i}}\, c^{i^{*}}_{m'_{i}}\ ;
& {\cal C}^{f}_{m_{f}m'_{f}} &= c^{f}_{m_{f}}\, c^{f^{*}}_{m'_{f}}\ , \nonumber \\
{\cal C}^{n}_{m_{n}m'_{n}}&= c^{n}_{m_{n}}\, c^{n^{*}}_{m'_{n}}\ ;
& {\cal C}^{n'}_{m_{n'}m'_{n'}} &= c^{n'}_{m_{n'}}\, c^{{n'}^{*}}_{m'_{n'}}\ .
\end{align}
Each of these matrices ${\cal C}^{i}, {\cal C}^{f}, {\cal C}^{n}, 
{\cal C}^{n'}$
are clearly hermitian and satisfy the condition
${\cal C}^{2} = {\cal C}$. Note that the ${\cal P}^{nn'}_{pp'}$ depend 
not only on $n$ and $n'$ but also on the direction 
$(\widetilde \theta_{B}, \widetilde \phi_{B})$ and the strength $B$ of the
magnetic field ${\vec B}$ and on 
$(A, \eta; \alpha_{Q}, \beta_{Q}, \gamma_{Q})$ characterising the
electric quadrupole field (because of 
Eq.~(\ref{ast}) for the $c^{k}_{m}$ with $k=i,f,n,n'$), apart from the angles
$(\theta', \phi')$ of the incident and $(\theta, \phi)$ of the scattered
radiation. Explicitly, therefore, ${\cal P}^{nn'}_{pp'} \equiv
{\cal P}^{nn'}_{pp'}(\theta,\phi; \theta'\phi'; {\vec B}; A, \eta;
\alpha_{Q},\beta_{Q},\gamma_{Q})$ for any given $J_{i}, J_{f}$.   
In the case of resonance scattering, when only a single intermediate
state $|\psi_{n}\rangle$ with $E_{n}=E_{i}+\omega'=E_{f}+\omega$
contributes to Eq.~(\ref{tmumu}), one can replace the summation over
$n,n'$ by $n=n'$ corresponding to a single excited level, whereas the
double summation over $n,n'$ has to be retained in Hanle scattering.

\section{Particular Case}
We consider the simple case of scattering with electric dipole transitions
between a total angular momentum zero lower level and a total angular momentum 
one upper level, i.e., $J_{l}=J_{i}=J_{f}=0$ and $J_{u}=J_{n}=1$. 
Clearly, ${\cal C}^{i} = {\cal C}^{f} = 1$. We may then use Eq.~(\ref{tme}) 
to simplify the product ${\cal E}^{\dag}(\mu'')\, {\cal E}(\mu)$ and use the 
complex conjugate of Eq.~(\ref{tme}) to simplify the product 
${\cal A}(\mu')\, {\cal A}^{\dag}(\mu''')$ in Eq.~(\ref{spm}), so that
\begin{align}
{\cal P}^{nn'}_{pp'} &= \frac{1}{2}\, g^{nn'}_{p}(\theta, \phi)\, 
g^{nn'}_{p'}(\theta', \phi')^{*}\ ; & n,n' &=1,2,3 \ ,
\end{align}
in terms of 
\begin{eqnarray}
& & g^{nn'}_{p}(\theta, \phi)= |{\cal J}_{1}|^{2} \sum^{2}_{\lambda=0}\sum^{1}_{m',m''=-1}
C(1,1,\lambda;m',-m'',m)\nonumber \\
& & (-1)^{m''} \sum^{\lambda}_{\xi=-\lambda} f_{p}(\lambda,\xi)\, 
D^{\lambda}_{m \xi}(\phi,\theta,0)\,
c^{n}_{m''}\, c^{{n'}^*}_{m'}\ ,
\end{eqnarray}
where
\begin{eqnarray}
f_{0}(\lambda,\xi) &=& \frac{2}{\sqrt{3}}\, \delta_{\xi,0} \Bigl[
\delta_{\lambda,0} + \frac{1}{\sqrt{2}}\, \delta_{\lambda,2} \Bigr]\ , \nonumber \\
f_{1}(\lambda,\xi) &=& - \delta_{\lambda,2} \Bigl[ \delta_{\xi,2} + \delta_{\xi,-2} \Bigr]\ ,
\nonumber \\
f_{2}(\lambda,\xi) &=& {\rm i}\, \delta_{\lambda,2} \Bigl[ \delta_{\xi,2} - \delta_{\xi,-2}\Bigr]\ ,
\nonumber \\
f_{3}(\lambda,\xi) &=& \sqrt{2}\, \delta_{\xi,0}\, \delta_{\lambda,1}\ .
\end{eqnarray}

In the absence of the electric field, $n=M, n'=M'$ and the $c^{M}_{m}=
D^{1}_{mM}(\phi_{B}, \theta_{B}, 0)$, 
leading to the well-known Hanle scattering phase matrix given by 
Eqs.~(9) to (16) of Landi and Landi~\citep{2landi} if $\Phi_{M} \Phi^{*}_{M'}$
can be assumed to be independent of $M,M'$ in the limiting case of weak fields. 
If the Doppler convolution is effected following
exactly the procedure outlined by Stenflo~(1998), the Hanle-Zeeman
scattering matrix represented by Eqs.~(49) and (50) of Stenflo~(1998)
is recovered for $\theta_{B}=\phi_{B}=0$. In the case of strong fields
i.e, if $g\, B$ is large compared to the line widths (Zeeman effect), one may set
$M=M'$ and recover Eq.~(52) of Stenflo~(1998) for $\theta_{B}=\phi_{B}=0$
and the results obtained much earlier by Obridko~\cite{obridko}.

\section{ Numerical Results and Discussion}
If we consider the simplest geometry of the combined magnetic and quadrupole
electric fields with ${\vec B}$ along the ${\rm Z}$-axis of the PAF with
the PAF itself coinciding with the astrophysical frame i.e., 
$\alpha_{Q}=\beta_{Q}=\gamma_{Q}=\widetilde \theta_{B}=\widetilde \phi_{B}=
\theta_{B}=\phi_{B}=0$, the upper level with 
$J_{u}=1$ is split into three levels $n=1,2,3$~\citep{Oo} with energies 
\begin{align}
E_{1} &= -2\, r\, g\, B \ ; & E_{2,3} & = 
 (r \mp s )g\, B\ ,
\label{eta}
\end{align}
where $r = A/gB$ the ratio of the electric quadrupole and magnetic
field strengths, $s=(r^{2}\eta^{2}+1)^{1/2}$ and the corresponding 
eigen states $|\psi_1\rangle, |\psi_2\rangle$ and $|\psi_3\rangle$
characterized by 
\begin{eqnarray}
c^{1}_{0}&=&1,\ \ \ c^{1}_{\pm 1}=c^{2,3}_{0}=0 \nonumber \\
c^{3}_{\pm 1} &=& \pm c^{2}_{\mp 1} = (s + r\, \eta \pm 1)/2(s^{2}+r\,
\eta\, s)^{1/2}\ ,
\end{eqnarray}
using Eqs.~(\ref{ast}) and (\ref{abg}), with the electric quadrupole
field strength $A > 0$.

\begin{figure}
\includegraphics[width=4.5cm,height=3.7cm]{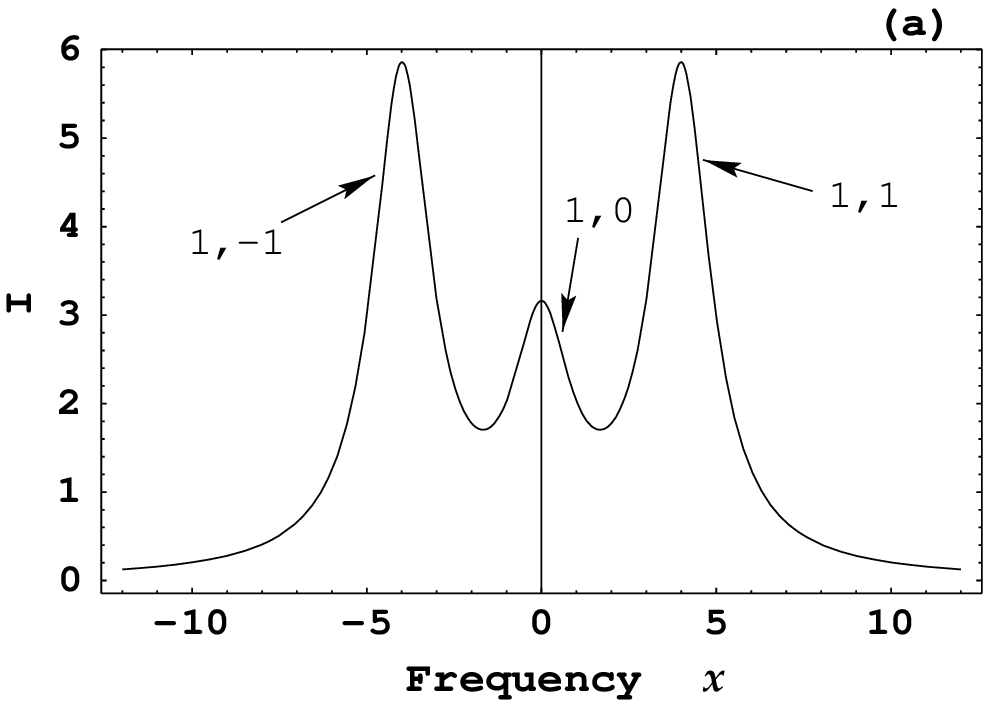}
\includegraphics[width=4.5cm,height=3.7cm]{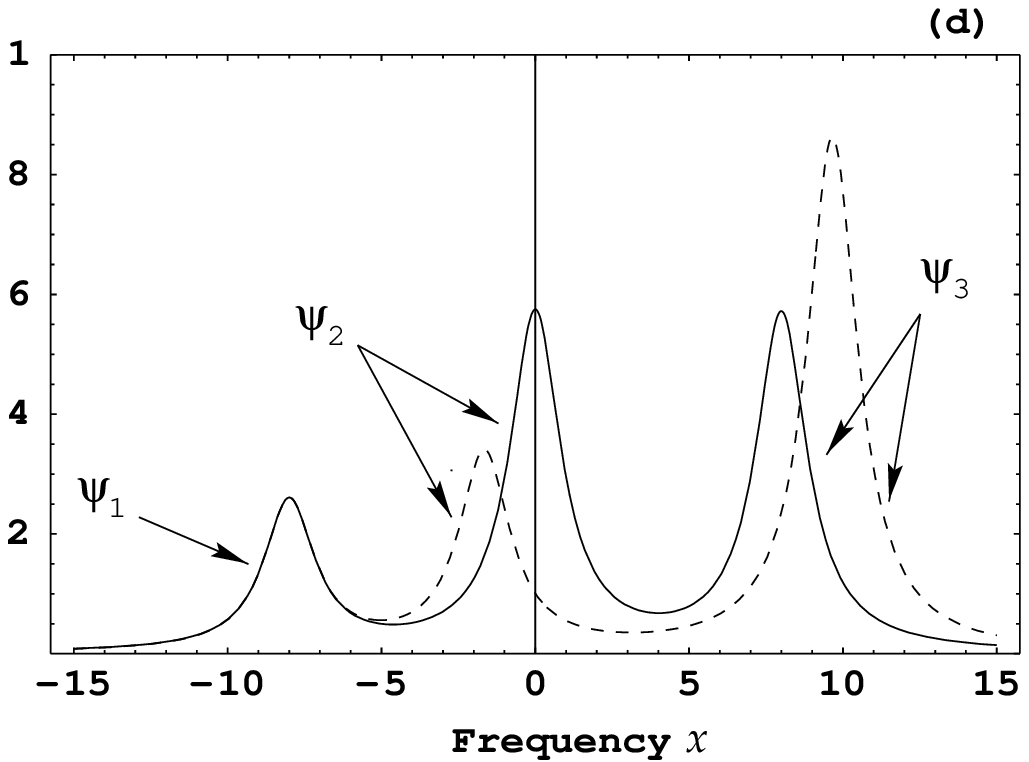}\\[.2cm]
\includegraphics[width=4.5cm,height=3.7cm]{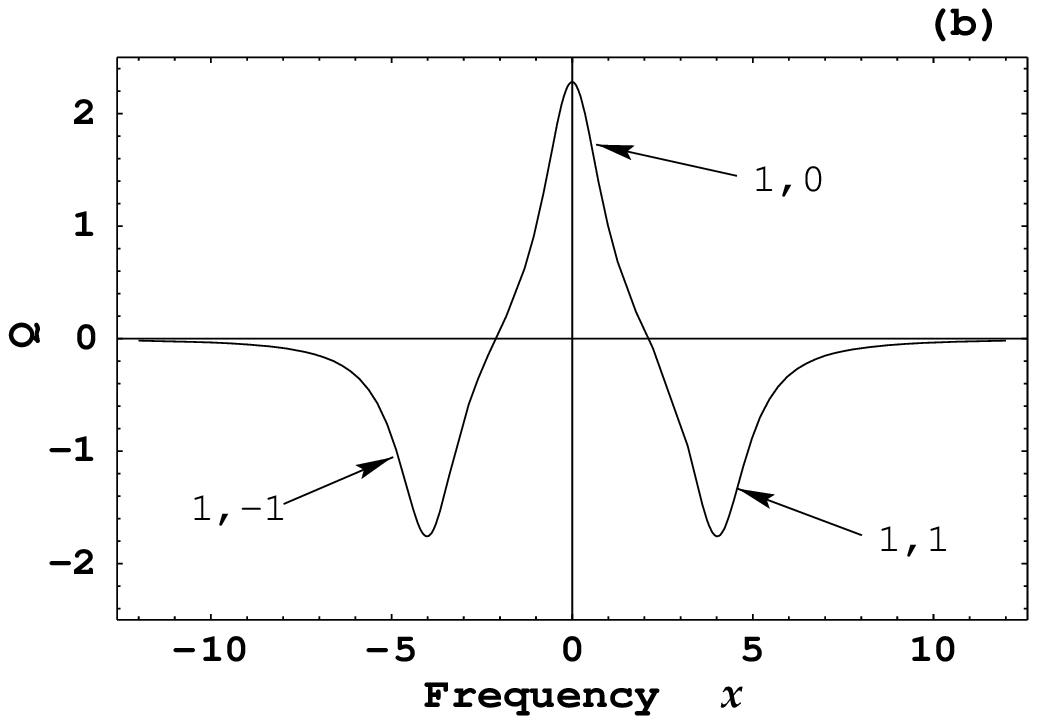}
\includegraphics[width=4.5cm,height=3.7cm]{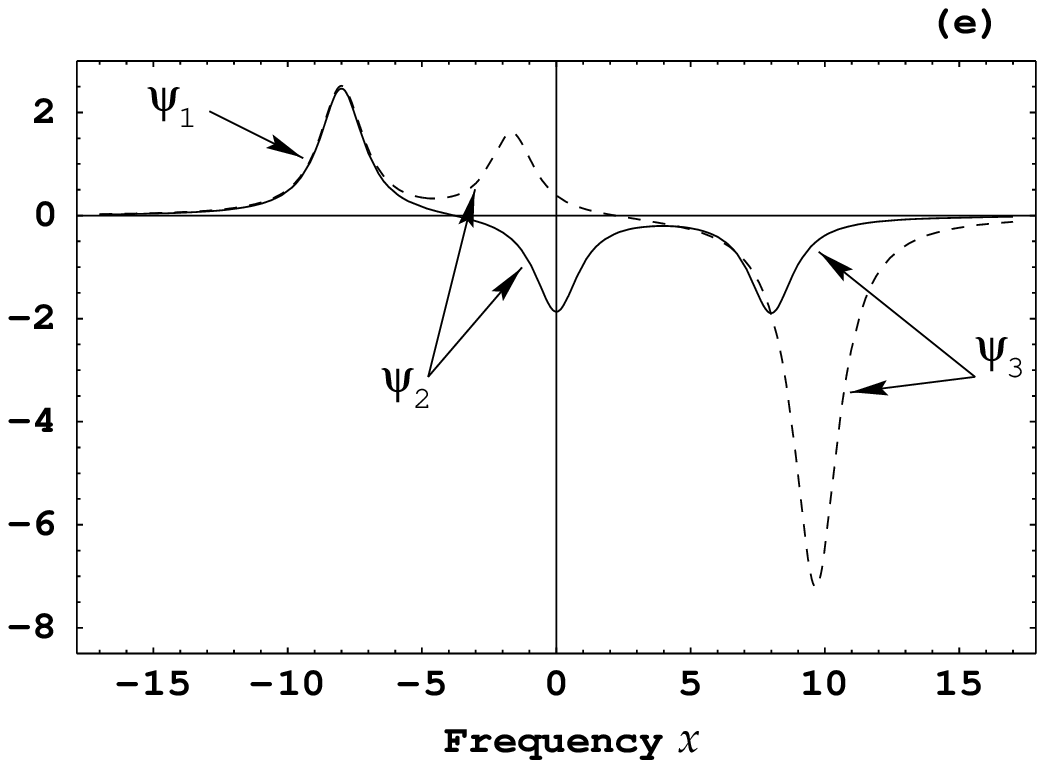}\\[.2cm] 
\includegraphics[width=4.5cm,height=3.7cm]{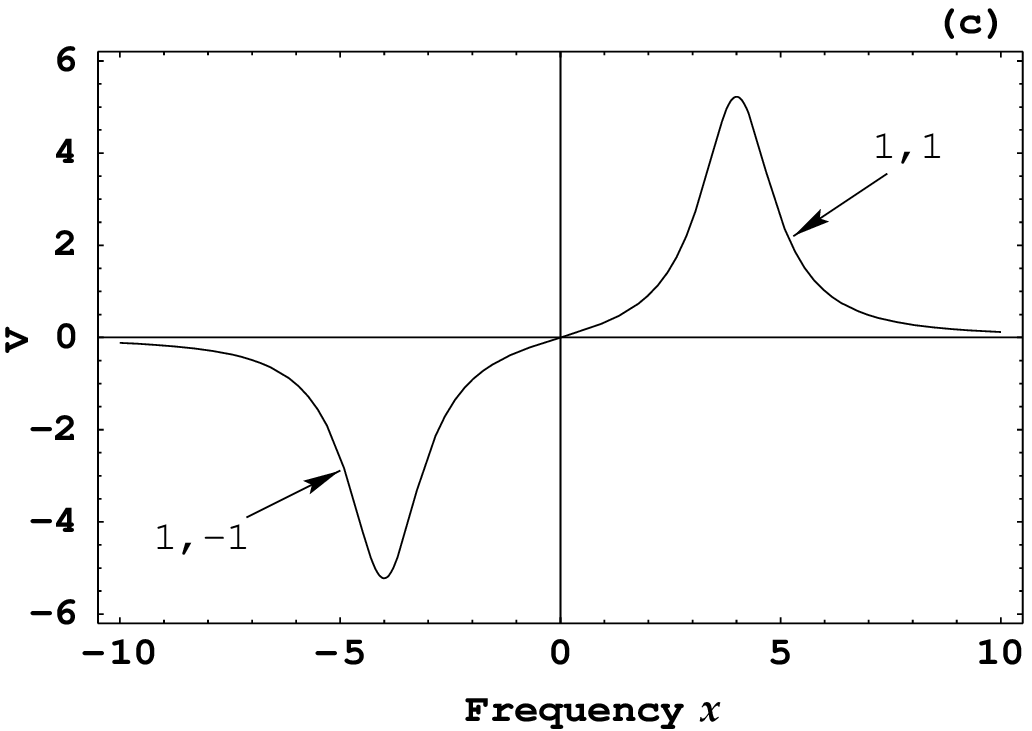}
\includegraphics[width=4.5cm,height=3.7cm]{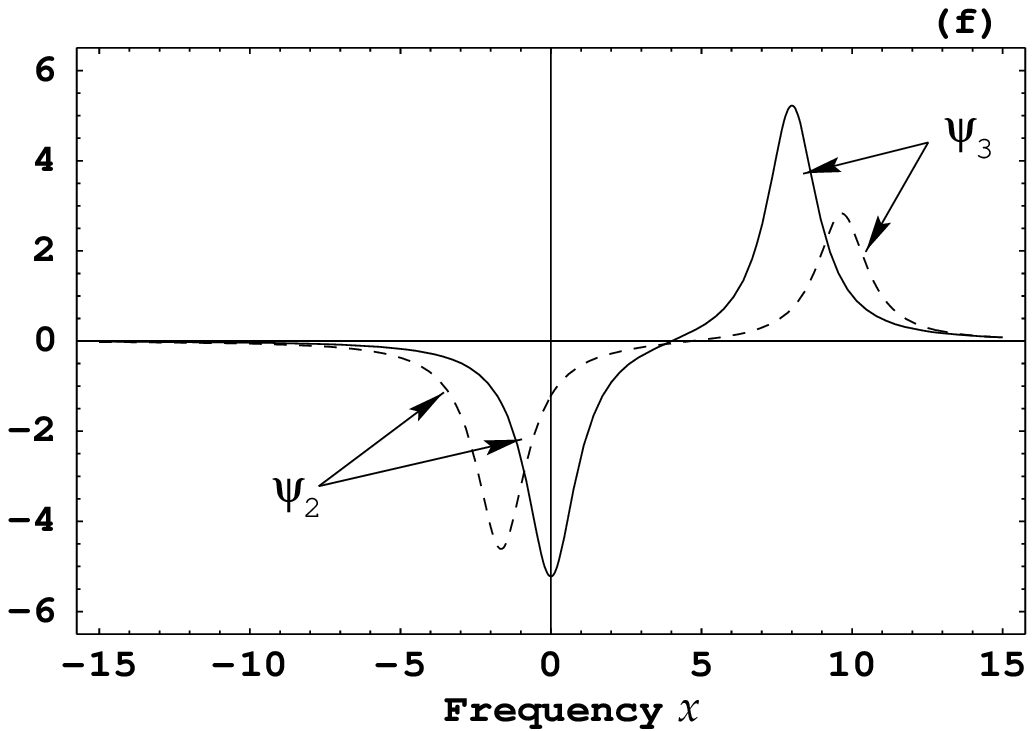}
\caption{ Stokes line profiles in arbitrary units comparing the Zeeman 
scattering in a pure magnetic field
(panels ${\bf a}-{\bf c}$), and scattering under the combined magnetic and electric 
quadrupole fields (panels ${\bf d}-{\bf f}$) with $r=1$. The solid lines correspond
to asymmetry parameter $\eta=0$, and the dashed lines to $\eta=1$. In all panels,
${\it x}$ is the frequency displacement from the line center in natural width units.
\label{plot} }
\end{figure}

To understand the combined effect of magnetic and electric quadrupole fields,
we present in Figs.~(\ref{plot}d-\ref{plot}f) the general behavior of the 
scattered Stokes line profiles ${\boldsymbol{\cal S}}$ for a given 
unpolarized incident radiation, 
${\boldsymbol{\cal S}'} =(1\  0\  0\  0)^{T}$,
for particular choices of the directions
$(\theta'=\frac{\pi}{4}, \phi'=0)$ and  
$(\theta=\frac{\pi}{4}, \phi=\pi)$.
We compare these with the pure Zeeman scattering case 
(see Figs.~\ref{plot}a-\ref{plot}c). In the Stokes $Q$ profile,
the positive maximum at the line center
and the negative maximum symmetrically placed at $\sigma_{\pm 1}$ components,
which are typical of the well known Zeeman effect. The maximum of the $V$ profile
at $\sigma_{\pm1}$ components have opposite sign, which is also a well known
characterisitc of Zeeman effect. 
We assume here that the magnetic field and the quadrupole electric field are equally
strong (i.e., $r= A/B = 1$). We also assume $B$ to be 
four times the natural line width and set $n=n'=1,2,3$. The solid and dashed curves in
Figs.~(\ref{plot}d-\ref{plot}f) are computed for the values of $\eta=0$
and $1$ respectively. In the combined fields case, the line component 
arising due to the $|\psi_{1}\rangle=|1,0\rangle$ state (which represents the 
central component in the corresponding pure Zeeman case) is positioned in the 
red wing (see Figs.~\ref{plot}d-\ref{plot}f). 
The unequal strengths of the scattered line profiles 
arising from $|\psi_{2}\rangle$ and $|\psi_{3}\rangle$ states are clearly seen in 
all the scattered Stokes line profiles $(I, Q, V)$. This is due to the 
weighted superposition of the magnetic substates $|1,-1\rangle$ and 
$|1,1\rangle$. In the scattered Stokes $V$ line profile, $|\psi_{1}\rangle$ state 
does not contribute, as in the case of pure Zeeman scattering.
Therefore the shape of the scattered Stokes $V$ profile is similar to the Zeeman case, 
except for the shifting of the position and the change in the relative strength 
of components when $\eta > 0$. If $\eta$ vanishes, the strengths of both 
components are same.

\acknowledgement
The authors are indebted to Prof.~J.~O.~Stenflo for kindly examining the 
manuscript, and for useful remarks, and  comments. They also wish to thank
an anonymous referee of this paper for pertinent remarks and suggestions that
proved very useful while revising the manuscript.
One of the authors (YYO) wishes to express her gratitude to the Chairman, Department of
Physics, Bangalore University and the Director, Indian Institute of Astrophysics (IIA),
for providing research facilities. She gratefully acknowledges the award of a
scholarship by Indian Council for
Cultural Relations (ICCR), Ministry of External Affairs (MEA), Government of India.
Another author (GR) is grateful to Professor B.~V.~Sreekantan,
Professor R.~Cowsik and Professor J.~H.~Sastri for much encouragement and facilities
provided for research at IIA. 

\end{document}